# Mantle Decompression Thermal-Tsunami

**J. Marvin Herndon**
**Transdyne Corporation**
**San Diego, CA 92131 USA**

**mherndon@san.rr.com**
**http://NuclearPlanet.com  http://UnderstandEarth.com**

**February 13, 2006**

**Abstract:** Previously in geophysics, only three heat transport processes have been considered: conduction, radiation, and convection or, more generally, buoyancy-driven mass transport. As a consequence of whole-Earth decompression dynamics, I add a fourth, called mantle decompression thermal-tsunami, which may emplace heat at the base of the crust from a heretofore unanticipated source.

It is known through experience in deep mines and with bore-holes that temperature increases with depth in the crust. For more than half a century geophysicists have made measurements of near-surface continental and oceanic heat flow with the aim of determining the Earth's heat loss. Pollack *et al.*[1] estimate a global heat loss of 44.2 TW (1 TW=$10^{12}$ W) based upon 24,774 observations at 20,201 sites.

Numerous attempts have been made to reconcile measured global heat loss with radionuclide heat production from various geophysical models closely involved with plate tectonics. Usually, models are contrived to yield the very result they model, but in this case there is a problem. Currently popular models rely upon radiogenic heat for geodynamic processes, geomagnetic field generation, and for the Earth's heat loss. The problem is that radionuclides cannot even satisfy just the global heat loss requirements.

Previous estimates of global heat production invariably come from the more-or-less general assumption that the Earth's current heat loss consists of the steady-state heat production from long-lived radionuclides ($^{235}$U, $^{238}$U, $^{232}$Th, and $^{40}$K). Estimates of present-day global radiogenic heat production, based upon chondritic abundances, typically range from 19 TW to 31 TW. These represent an upper limit through the tacit assumption of rapid heat transport irrespective of assumed radionuclide locations. The short-fall in heat production, relative to Earth's measured heat loss[1], has led to speculation that the difference might be accounted for by residual heat from Earth's formation $4.5 \times 10^9$ years ago, ancient radiogenic heat from a time of greater heat production, or, perhaps, from a yet unidentified heat source[2].

The purpose of this brief communication is to disclose a heretofore unanticipated heat transport mechanism and heat source capable of emplacing heat at the mantle-crust-interface at the base of the crust.

The principal consequences of Earth's formation from within a giant gaseous protoplanet are profound and affect virtually all areas of geophysics in major, fundamental ways[3]. Principal implications result (*i*) from Earth having been compressed by about 300 Earth-masses of primordial gases which provides a major source of energy for geodynamic processes, and (*ii*) from the deep-interior having a highly-reduced state of oxidation which results in great quantities of uranium and thorium existing within the Earth's core, and leads to the feasibility of the georeactor, a hypothesized natural, nuclear fission reactor at the center of the Earth as the energy source for the geomagnetic field[4-7]. These consequences have led to a different way of envisioning geodynamics, recently published in *Current Science*[8], called *whole-Earth decompression dynamics*.

Formation of the Earth as the rock-plus-alloy kernel of a giant gaseous Jupiter-like planet, as I have shown[3,9,10], leads to the Earth as we know it being compressed to about 64% of its present diameter, and having a contiguous uniform shell of continental matter covering its rocky surface. After being stripped of its great, Jupiter-like overburden of volatile protoplanetary constituents, presumably by the high temperatures and/or by the violent activity, such as T Tauri-phase solar wind[11-13], associated with the thermonuclear ignition of the Sun, the Earth would inevitably begin to decompress, to rebound toward a new hydrostatic equilibrium. The initial whole-Earth decompression is expected to result in a global system of major *primary* cracks appearing in the rigid crust which persist and are identified as the global, mid-oceanic ridge system, just as explained by Earth expansion theory. But here the similarity with that theory ends. Whole-Earth decompression dynamics sets forth a different mechanism for whole-Earth dynamics which involves the formation of *secondary* decompression cracks and the in-filling of those cracks, a process which is not limited to the last 200 million years, the maximum age of the seafloor.

As the Earth subsequently decompresses and swells from within, the deep interior shells may be expected to adjust to changes in radius and curvature by plastic deformation. As the Earth decompresses, the area of the Earth's rigid surface increases by the formation of secondary decompression cracks often located near the continental margins and presently identified as submarine trenches. These secondary decompression cracks are subsequently in-filled with basalt, extruded from the mid-oceanic ridges, which traverses the ocean floor by gravitational creep, ultimately plunging into secondary decompression cracks, thus emulating subduction, but without necessitating mantle convection.

One of the consequences of Earth formation as a giant, gaseous, Jupiter-like planet[9], as described by whole-Earth decompression dynamics[8,10,14], is the existence of a vast reservoir of energy, the stored energy of protoplanetary compression, available for driving geodynamic processes related to whole-Earth decompression. Some of that energy, I submit, is emplaced as heat at the mantle-crust-interface at the base of the crust through the process of *mantle decompression thermal-tsunami*. Moreover, some radionuclide heat may not necessarily contribute directly to crustal heating, but rather to replace the lost heat of protoplanetary compression, which helps to facilitate mantle decompression.

Previously in geophysics, only three heat transport processes have been considered: conduction, radiation, and convection or, more generally, buoyancy-driven mass transport. As a consequence of whole-Earth decompression dynamics, I add a fourth, called mantle decompression thermal-tsunami.

As the Earth decompresses, heat must be supplied to replace the lost heat of protoplanetary compression. Otherwise, decompression would lower the temperature, which would impede the decompression process.

Heat generated within the core from actinide decay and/or fission[15] may enhance mantle decompression by replacing the lost heat of protoplanetary compression. The resulting decompression, beginning at the bottom of the mantle, will tend to propagate throughout the mantle, like a tsunami, until it reaches the impediment posed by the base of the crust. There, crustal rigidity opposes continued decompression, pressure builds and compresses matter at the mantle-crust-interface, resulting in compression heating. Ultimately, pressure is released at the surface through volcanism and through secondary decompression crack formation and/or enlargement.

Mantle decompression thermal-tsunami, as outlined above, poses a new explanation for a portion of the internal heat being lost from the Earth. It may prove as well to be a significant energy source for earthquakes and volcanism, as these geodynamic processes appear concentrated along secondary decompression cracks.